\documentclass[fleqn,usenatbib]{rasti}

\usepackage{newtxtext,newtxmath}

\usepackage[T1]{fontenc}

\DeclareRobustCommand{\VAN}[3]{#2}
\let\VANthebibliography\thebibliography
\def\thebibliography{\DeclareRobustCommand{\VAN}[3]{##3}\VANthebibliography}

\usepackage{graphicx}	
\usepackage{amsmath}	
\usepackage{physics}
\usepackage{orcidlink}

\newcommand{\Pwd}{P_\mathrm{WD}}
\newcommand\xb{\vb{x}}
\newcommand\yb{\vb{y}}

\title[
    Polluted white dwarfs from \textit{Gaia} XP spectra
]{
    Finding rare classes in large datasets: the case of polluted white dwarfs from \textit{Gaia} XP spectra
}

\author[
    X. Byrne et al.
]{
    Xander Byrne
    \orcidlink{0000-0001-9488-238X},$^{1}$\thanks{E-mail: \href{mailto:xbyrne@ast.cam.ac.uk}{xbyrne@ast.cam.ac.uk}}
    Amy Bonsor
    \orcidlink{0000-0002-8070-1901},$^{1}$
    Laura K. Rogers
    \orcidlink{0000-0002-3553-9474},$^{2,1}$
    Mariona Badenas-Agusti$^{1}$
    \orcidlink{0000-0003-4903-567X}
\\
$^{1}$Institute of Astronomy, University of Cambridge, Madingley Road, Cambridge CB3 0HA, UK\\
$^{2}$NOIRLab, 950 N Cherry Ave, Tucson, AZ, 85719, USA\\
}

\date{Accepted XXX. Received YYY; in original form ZZZ}

\pubyear{\the\year{}}

\begin{document}
\label{firstpage}
\pagerange{\pageref{firstpage}--\pageref{lastpage}}
\maketitle

\begin{abstract}
The \textit{Gaia} mission's third data release recorded low-resolution spectra for about $100\,000$ white dwarf candidates.
A small subset of these spectra show evidence of characteristic broad Ca~\textsc{ii} absorption features, implying the accretion of rocky material by so-called polluted white dwarfs -- important probes of the composition of exoplanetary material.
Several supervised and unsupervised data-intensive methods have recently been applied to identify polluted white dwarfs from the \textit{Gaia} spectra.
We present a comparison of these methods, along with the first application of $t$-distributed stochastic neighbour embedding ($t$SNE) to this dataset.
We find that $t$SNE outperforms the similar technique Uniform Manifold Approximation and Projection (UMAP), isolating over 50\% more high-confidence polluted candidates, including 39 new candidates which are not selected by any other method investigated and which have not been observed at higher resolution.
Supervised methods benefit greatly from data labels provided by earlier works, selecting many known polluted white dwarfs which are missed by unsupervised methods.
Our work provides a useful case study in the selection of members of rare classes from a large, sporadically labelled dataset, with applications across astronomy.
\end{abstract}




\section{Introduction}

White dwarfs (WDs) are remnant stellar cores, the final form of all main-sequence (MS) stars with zero-age masses below $9$--$12\mathrm{M}_{\sun}$ \citep{althaus10, althaus21, lauffer18}.
WDs constitute a highly uniform population, and are therefore ideal astrophysical laboratories with which to probe a wide range of astrophysical phenomena, including the bulk composition of exoplanetary material.

The presence of metal `pollution' in the spectra of WDs reveals the composition of rocky bodies in the system.
Following a star's post-MS evolution, exoplanetary bodies can be scattered onto star-grazing orbits \citep{debes02, bonsor11, frewen14, mustill18, maldonado20dyn, maldonado20und}, tidally disrupted \citep{jura03, veras14}, and then accreted onto the WD \citep{brouwers22}.
This has been shown to be the dominant mechanism behind the appearance of metal features in the spectra of so-called polluted WDs \citep[e.g.,][]{farihi10, veras16}.
By fitting WD atmosphere models to the polluted WD spectra \citep[e.g.][]{koester05, koester11, dufour07, hollands17, hollands18, blouin18, badenasagusti24}, one can post-mortem constrain the bulk composition, and thus geology, of the accreted parent body \citep[e.g.][]{xu21}.

The \textit{Gaia} mission \citep{gaia} has revolutionized the study of WDs.
The mission's second data release (DR2) identified over $262\,000$ high-confidence WD candidates, an increase by almost an order of magnitude on the then-state-of-the-art \citep{gentilefusillo19, gaiadr2}.
This increased further to over $359\,000$ thanks to the improved astrometry and photometry of \textit{Gaia}'s Early Data Release 3 \citep[EDR3;][]{gentilefusillo21, gaiadr3}.
In addition to astrometric and photometric measurements, the full \textit{Gaia} DR3 also provides low-resolution ($R\sim70$) spectra for 219 million sources
\citep{carrasco21, gaiadr3}, including $\approx100\,000$ WD candidates \citep{gentilefusillo21}.
These spectra -- called `BP/RP spectra'; collectively `XP spectra' -- are stored not as fluxes in a sequence of wavelength bins, but as a pair of 55-coefficient Hermite polynomials.
They can therefore be represented by a 110-dimensional coefficient vector $\xb\in\mathbb{R}^{110}$.
The Hermite polynomials approximating the spectrum can be reconstructed using the \texttt{GaiaXPy} Python package\footnote{
    \url{https://gaia-dpci.github.io/GaiaXPy-website/}
}.

While an XP spectrum is not sufficient to classify a WD as polluted or not (let alone fit elemental abundances) they indicate which WDs from this large dataset to observe with targeted spectroscopic campaigns at higher resolution.
The high surface gravities of WDs ($\sim 10^8\,\text{g cm}^{-2}$) cause significant collisional broadening, allowing metal pollution to be detected in even low-resolution spectra, usually from the strong Ca H and K lines at $3933$ and $3968~\text{\AA}$.

Several techniques have recently been applied to the large \textit{Gaia} XP sample to identify polluted WD candidates:
\begin{itemize}
\item \textbf{Uniform manifold approximation and projection (UMAP; \citealt{mcinnes18})}.
UMAP is an example of a dimensionality reduction technique, wherein a dataset of high-dimensional points is projected into a two-dimensional map, while approximately preserving the original structure of the dataset.
In the case of UMAP, this is achieved by finding a 2D submanifold of the high-dimensional data space that is as close as possible to the data points. 
\citet{kao24} apply UMAP to a sample of $96~134$ XP spectra, identifying a distinct group of 465 WDs of which 90 were known \textit{a priori} to be polluted.
The remaining 375 are subject to an ongoing high-resolution spectroscopic campaign; they report that 99 per cent of those observed at the time of publication showed multiple metal lines.
\item \textbf{Self-organizing maps (SOMs; \citealt{kohonen82})}.
SOMs employ neural networks to fit a flexible grid of neurons to the data, `assigning' each data point to a particular neuron together with similar data points.
This partitions the dataset, such that within each subset the data are similar to each other.
\citet{perezcouto24} use a sequence of two SOMs: one to remove contaminants such as quasars and galaxies, and another to partition the dataset into spectral classes.
They find two populations of 249 and 218 polluted WD candidates, the latter of which was not identified by UMAP.
\item \textbf{Gradient tree boosting \citep{friedman01}}.
Gradient tree boosting trains a sequence of decision trees to regress the error on the prediction from the tree before it.
\citet{vincent24} apply gradient tree boosting to classify 101~783 WD candidates into one of six primary WD spectral types:
DA (characterised by hydrogen features in the spectrum),
DB (neutral helium features),
DC (no features),
DO (ionized helium),
DQ (carbon),
and DZ (metals)\footnote{
    This classification scheme was first outlined in \citet{sion83}; the `D' stands for `degenerate'.
}.
They classify $1~272$ WDs as DZs.
\item \textbf{Random forests \citep{breiman01}}.
A random forest consists of a large ensemble of decision trees, which are trained to classify data.
\citet{garciazamora25} apply this method to classify $78\,920$
\textit{Gaia} WDs within $500\,\text{pc}$, to (i) differentiate between DA and non-DA WDs; (ii) classify non-DA WDs into DB, DC, DQ, and DZ WDs.
They classify 785 WDs as DZ.
\end{itemize}

The first two of these methods are \textit{unsupervised}; the latter two are \textit{supervised}.
Supervised methods rely on a labelled training set, which in this case means WDs in the sample that have a known spectral classification, perhaps from existing higher-resolution observations.
Although data labels constitute highly relevant information, any biases or imbalances present in the training data can propagate through to the resulting model's predictions.
Unsupervised methods do not require this ground-truth information, simply processing all of the data at face value.
They therefore evade label bias while enabling the serendipitous discovery of anomalies \citep[e.g.,][]{giles19, webb20}.

$t$-distributed Stochastic Neighbour Embedding ($t$SNE; \citealt{vandermaaten08}) is another example of an unsupervised method.
It is a dimensionality reduction technique, similar to UMAP, which attempts to project the dataset into 2D by preserving as well as possible the ``similarity'' between pairs of data points (see Section~\ref{sec:tsne}).
A key empirical difference between $t$SNE and UMAP is that UMAP prioritises the global structure of a dataset more than $t$SNE \citep{mcinnes18, fotopoulou24}.
The embeddings generated by $t$SNE may therefore better reveal the similarities between similar data points, at the cost of presenting a less realistic picture of the dataset at large.

\citet{steinhardt20} use $t$SNE to identify quiescent galaxies from UltraVISTA photometry \citep{mccracken12}.
\citet{hawkins21} likewise use $t$SNE to identify 416 metal-poor stars from a sample of $14\,000$ stars in the Hobby–Eberly Telescope Dark Energy Experiment (HETDEX; \citealt{gebhardt21}) survey.
More recently, \citet{byrne24c} demonstrate the use of $t$SNE to explore the intermediate-resolution Dark Energy Spectroscopic Instrument's Early Data Release (DESI EDR; \citealt{desiedr}).
Among their findings, cataclysmic variables (CVs) are identified from the catalogue at human-level recall in seconds.

This work compares the ability of various techniques, including $t$SNE, to identify polluted WDs from \textit{Gaia} XP spectra.
The isolation of rare classes from a large dataset with sparse labels is a common problem in Astronomy, so our findings are expected to apply more broadly.
Section~\ref{sec:methods} outlines details on $t$SNE and characterizes the dataset.
Section~\ref{sec:results} demonstrates $t$SNE's ability to select polluted WDs, as well as other patterns that appear in the resulting embedding.
Section~\ref{sec:discussion} compares $t$SNE's capability in this task to that of other methods, and suggests use cases for the various techniques in attaining other insights from large datasets.
Section~\ref{sec:conclusion} concludes our work.

\section{Methods} \label{sec:methods}

\subsection{\textit{t}SNE} \label{sec:tsne}

$t$SNE is a technique for mapping a dataset of high-dimensional points $\xb\in\mathbb{R}^D$ into lower-dimensional points $\yb\in\mathbb{R}^d$, while preserving local structure probabilistically.
The lower dimensionality $d<D$ is usually either 2 or 3; we use $d=2$ throughout.
This dimensionality reduction is achieved by defining two \textit{similarity} functions -- one for the original $D$-dimensional space, another for $d$ dimensions -- and then minimizing the difference between the two similarity distributions with respect to the projected ($d$-dimensional) points.

The similarity between two points $\xb_i, \xb_j \in \mathbb{R}^D$ is defined by
\begin{equation}
p_{ij} = \frac{1}{2N} \qty(p_{i|j} + p_{j|i}),
\end{equation}
where
\begin{equation}
p_{i|j} = \frac{
    \exp\qty(-\norm{\xb_i - \xb_j} / 2\sigma^2)
}{
    \sum_{k\neq j} \exp\qty(-\norm{\xb_k - \xb_j} / 2 \sigma^2)
}.
\end{equation}
The parameter $\sigma$ is internally optimised such that this similarity distribution achieves a particular user-defined value of the \textit{perplexity} $\mathrm{Perp}(p_i)$:
\begin{equation}
\log_2\mathrm{Perp}(p_i) \equiv
    -\sum_j p_{j|i} \log_2 p_{j|i};
\end{equation}
the right-hand side being recognizable as the Shannon entropy of the similarity distribution in bits \citep{vandermaaten08}.
In 2D, the similarity is instead defined according to a Cauchy distribution:
\begin{equation}
q_{ij} = \frac{
    \qty(1 + \norm{\yb_i - \yb_j}^2)^{-1}
}{
    \sum_{k\neq j} \qty(1 + \norm{\yb_k - \yb_j}^2)^{-1}
}.
\end{equation}
The use of a Cauchy distribution -- which has broader tails than a normal distribution -- addresses the `crowding problem', a phenomenon permitting high-dimensional points to have many more close neighbours than 2D points \citep{vandermaaten08}.

Ideally, the pairwise distances $\norm{\yb_i-\yb_j}$ in the embedding should be as close as possible to the corresponding $\norm{\xb_i-\xb_j}$ in the original data space.
$t$SNE therefore proceeds by probabilistically minimizing the difference between the two similarity distributions, as quantified by the Kullback-Leibler divergence \citep{kullbackleibler}:
\begin{equation}
\mathcal{KL}(p||q)
\equiv \sum_{i, j} p_{ij} \log \qty(
    \frac{p_{ij}}{q_{ij}}
).
\end{equation}
The embedding is constructed by minimizing this divergence with respect to the 2D points $\{\yb_i\}$, for instance by gradient descent.

For our purposes, each vector $\xb_i$ is a 110-dimensional vector, each of whose components is an XP spectral coefficient.
To account for sources being of different apparent magnitude -- a parameter irrelevant to their classification -- these coefficients are first divided by the \textit{G}-band flux; we found that this normalization enabled easier isolation of polluted WD candidates in the embedding, compared to L2 or unit-Gaussian normalization.
Our results are not sensitive to perplexity values between about 30 and 120; we henceforth use a perplexity of 50.
We use the implementation of $t$SNE provided in the \textsc{scikit-learn} Python package \citep{sklearn} throughout, and use their default values for hyperparameters other than the perplexity.

\subsection{Sample selection}
The sample of $359\,073$ high-confidence WD candidates created by \citet{gentilefusillo21} was selected based on \textit{Gaia} EDR3 photometry, as well as a series of quality filters to exclude sources with e.g.\ unreliable astrometry.
Of these, $107\,797$ sources had XP spectra released in DR3, as obtained from the Gaia@AIP service\footnote{
    \url{https://gaia.aip.de/}
}.

Some of these sources inevitably contain low-signal-to-noise data, and further selection criteria are needed.
There seems to be little consensus as to how best to achieve this: previous works have performed various combinations of cuts on the distance, astrometric noise, number of \textit{Gaia} observations, or the \textit{probability of being a white dwarf} ($\Pwd$; \citealt{gentilefusillo15}).
Comparing these criteria, we suggest that the criteria of \citet{perezcouto24} are the most inclusive and well-justified.
These criteria are:
\begin{itemize}
\item $(\texttt{phot\_bp\_n\_obs} > 10)$ and $(\texttt{phot\_rp\_n\_obs} > 15)$, as recommended by \citet{andrae23};
\item $\texttt{visibility\_periods\_used} > 10$, as recommended by \citet{lindegren18}.
\end{itemize}
We note that \citet{perezcouto24} also list a cut on the \texttt{phot\_bp\_rp\_excess\_factor\_corrected}, however this same cut is already made in assembling the supersample of \citet[][their equation 21]{gentilefusillo21}, and is therefore not necessary.
Following these cuts, we obtain a sample of size $107\,164$.

\subsection{Evaluation of polluted WD selection} \label{sec:evaluation}

Each method for identifying polluted WDs will be imperfect: some polluted WDs will inevitably escape selection, and some other objects will be erroneously selected.
In evaluating the different methods, it is thus crucial to know which sources are genuinely polluted WDs.

Unambiguously identifying a polluted WD usually requires much higher-resolution spectroscopy than the \textit{Gaia} XP spectra, with an $R$ of at least $1~000$ and ideally $R>20~000$ to detect weaker pollution lines.
Such observations are however time- and resource-intensive and have therefore only been conducted for a small fraction of the dataset.
For most of the sources, their pollution status is unknown.
For those that \textit{have} been observed, they have either been confirmed as polluted WDs, or as WDs of different, non-polluted spectral type.

We categorize each source as either `known polluted', `known non-polluted' and `unknown' according to their classifications in three datasets, each of which collates higher-resolution observations.
These are:
\begin{itemize}
\item Montreal White Dwarf Database \citep[MWDD;][]{dufour17};
\item \textit{Gaia}-SDSS spectroscopic sample \citep{gentilefusillo21};
\item Planetary Enriched White Dwarf Database \citep[PEWDD;][]{williams24}.
\end{itemize}
If a source is assigned a `polluted' spectral class (DZ, DAZ etc.) in any of these datasets, we categorise it as `known polluted'.
If assigned a non-polluted spectral class in any dataset (e.g.\ DA), it is categorised as `known non-polluted'.
Otherwise, its pollution status is categorised `unknown'.

For WDs with weak metal lines, these features may be fundamentally undetectable in the low-resolution XP spectra.
Furthermore, some WDs have only been identified as polluted in the ultraviolet \citep[$\lambda < 300\,\text{nm}$; e.g.][]{koester14, ouldrouis24}, which is outside the XP spectral window.
As such, even an idealised method would not be able to identify all `known polluted' WDs.

When evaluating the performance of \textit{supervised} methods, it is important to note that many `known polluted' WDs likely appear in their training sets.
Supervised methods are much more likely to correctly classify data present in their training sets, so a large proportion of `known polluted' WDs will inevitably be identified by these methods.
However, it is unlikely that supervised methods would perform equally well on unseen data: as such, a lower proportion of the `unknown' WDs selected by these methods would be expected to be true polluted WDs, than the proportion of correctly classified \textit{known} polluted WDs.
A second reason for this is that WDs with an `unknown' pollution status are generally fainter (being less likely to have been observed at higher resolution).
Their XP spectra are thus on average lower-signal-to-noise than those used in the training set, which may make the supervised methods (inevitably trained on higher-signal-to-noise data) less likely to identify polluted WDs that have not been observed at higher resolution.

\section{Results} \label{sec:results}

\subsection{Identification of polluted WDs by \textit{t}SNE} \label{sec:results/tsne}

\begin{figure}
\centering
\includegraphics[width=\columnwidth]{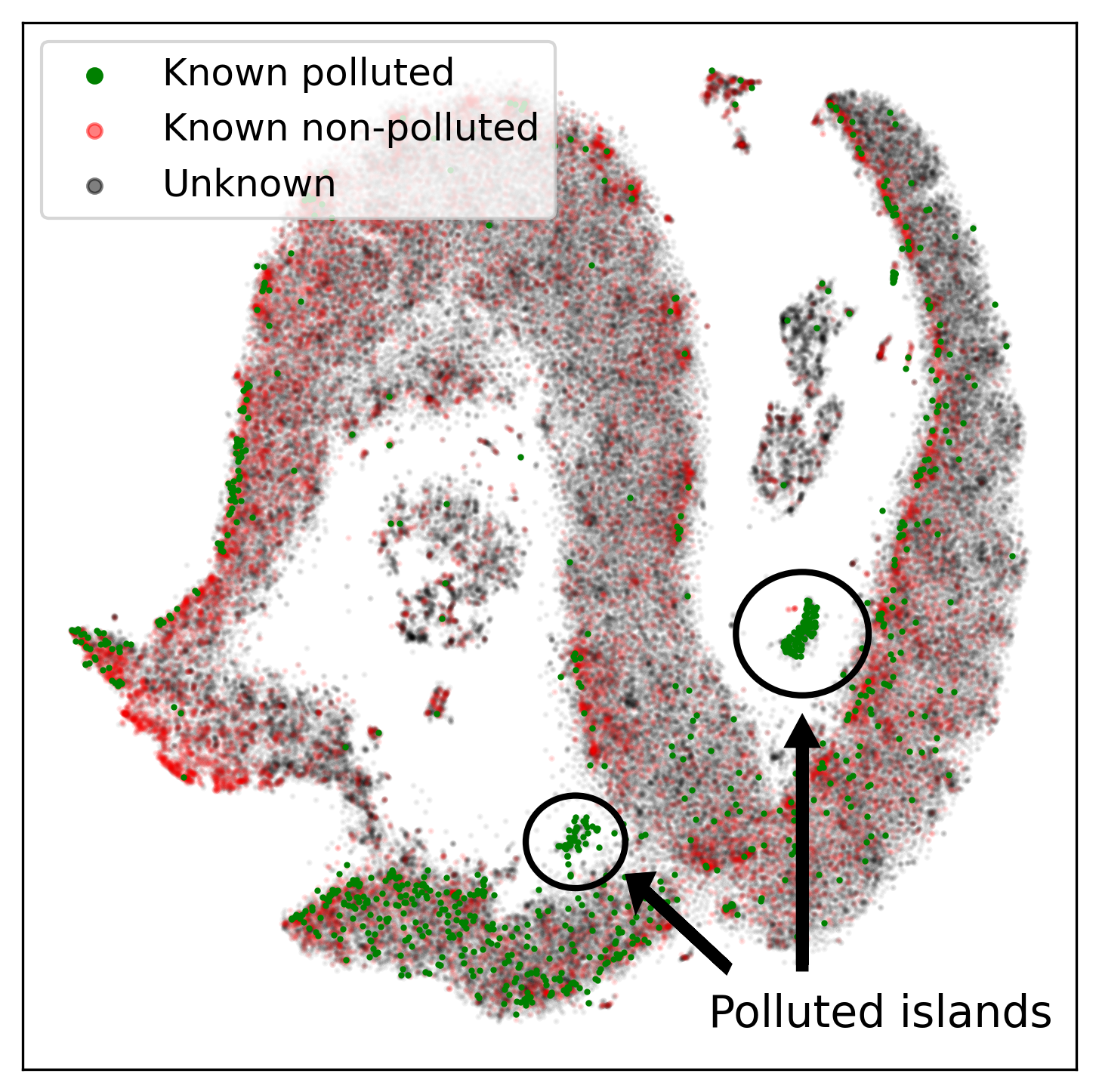}
\caption{
    $t$SNE embedding of normalized \textit{Gaia} WD candidate XP spectra.
    The embedding's primary feature is an `N'-shaped sequence, with a number of islands scattered around it.
    Two highlighted islands contain a large number of known polluted WDs, and few known non-polluted WDs.
}
\label{fig:tsneembedding}
\end{figure}

The \textit{t}SNE embedding of the sample (Fig.~\ref{fig:tsneembedding}) features a large `N'-shaped structure, surrounded by several smaller `islands'.
Two of these islands show a high purity of known polluted WDs, suggesting that many of the `unknown' WDs in these islands might be polluted WDs that have not yet been observed at higher resolution.

\begin{figure*}
\centering
\includegraphics[width=\textwidth]{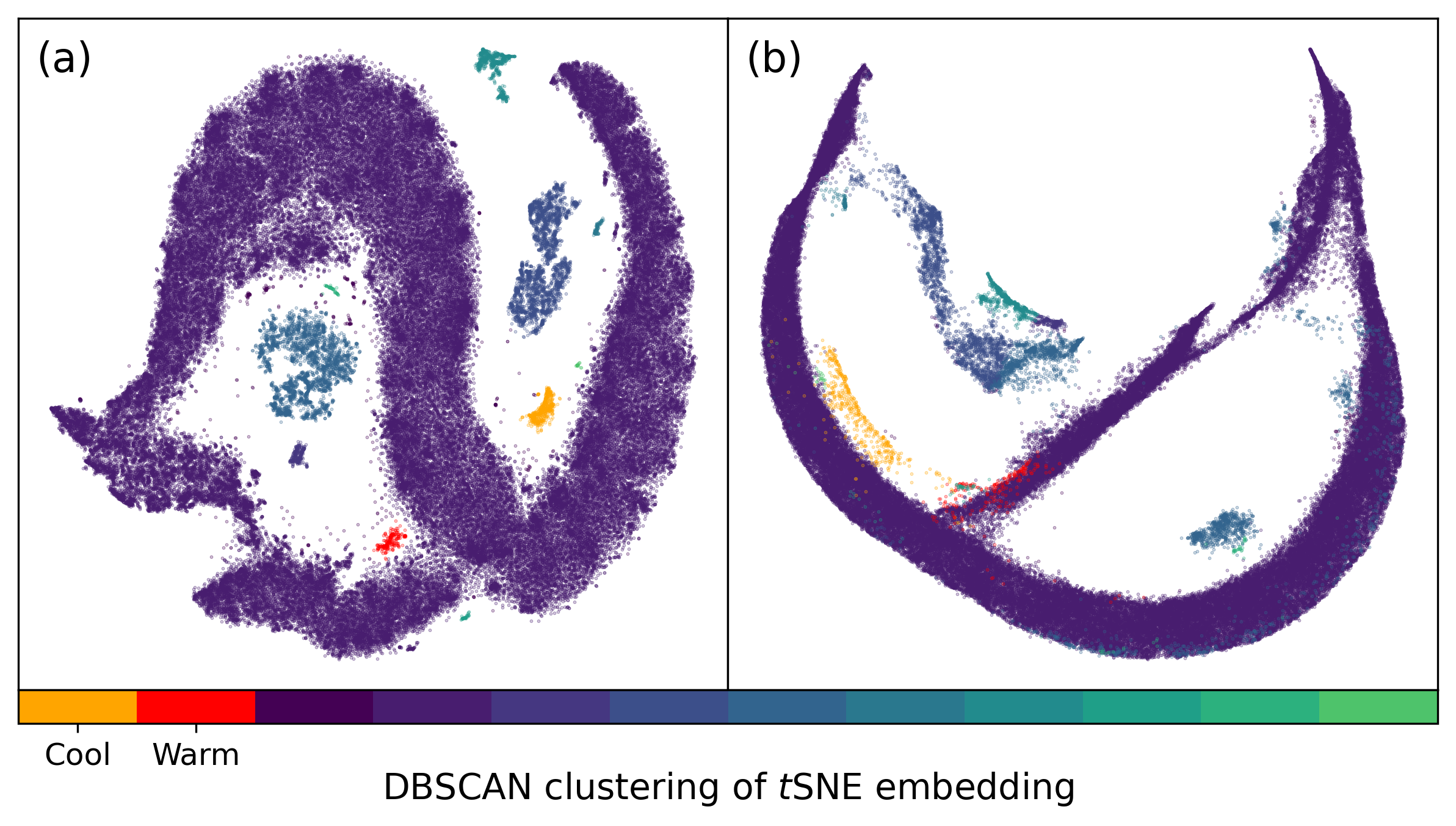}
\caption{
    (a) DBSCAN clustering applied to the $t$SNE embedding.
    Different clusters in the embedding are colour-coded with different colours; the two polluted islands (`cool' and `warm') are highlighted.
    (b) UMAP embedding of the sample, using the colour-coding from (a).
    The embedding is similar to that of \citet{kao24} (see their fig.~2), with the WDs from our `cool' island corresponding roughly to those found in their work.
    However, the WDs from our `warm' island are buried within the diagonal feature of the UMAP embedding.
}
\label{fig:tsneclustering}
\end{figure*}

Which sources belong to these islands can be decided objectively using clustering algorithms, such as the well-known Density-Based Spatial Clustering of Applications with Noise \citep[DBSCAN;][]{ester96} algorithm.
We apply the \texttt{scikit-learn} implementation of this algorithm to the $t$SNE embedding, with parameters\footnote{
    The two parameters \texttt{eps} and \texttt{min\_samples} define the clustering.
    Briefly, each member of a cluster either has at least \texttt{min\_samples} points within a distance \texttt{eps}, or is at most \texttt{eps} away from such a point.
} $\mathtt{eps}=2$ and $\mathtt{min\_samples}=30$, identifying twelve clusters, of which two correspond to the polluted islands (Fig.~\ref{fig:tsneclustering}(a)).
We refer to these as the `cool' and the `warm' islands, owing to their respective temperature distributions (see Section~\ref{sec:whytwo}).

The UMAP embedding constructed in \citet{kao24} also contained a feature populated mostly by many known polluted WDs (see their fig.~2 and 3), but it is slightly blurred with other features, making membership less certain for some sources.
However, the $t$SNE embedding contains sufficiently distinct clusters that algorithms such as DBSCAN can evaluate cluster membership objectively.
The reasons for this difference in clustering behaviour between the two algorithms are discussed in Section~\ref{sec:tsnevumap}.

\begin{figure*}
\centering
\includegraphics[width=\textwidth]{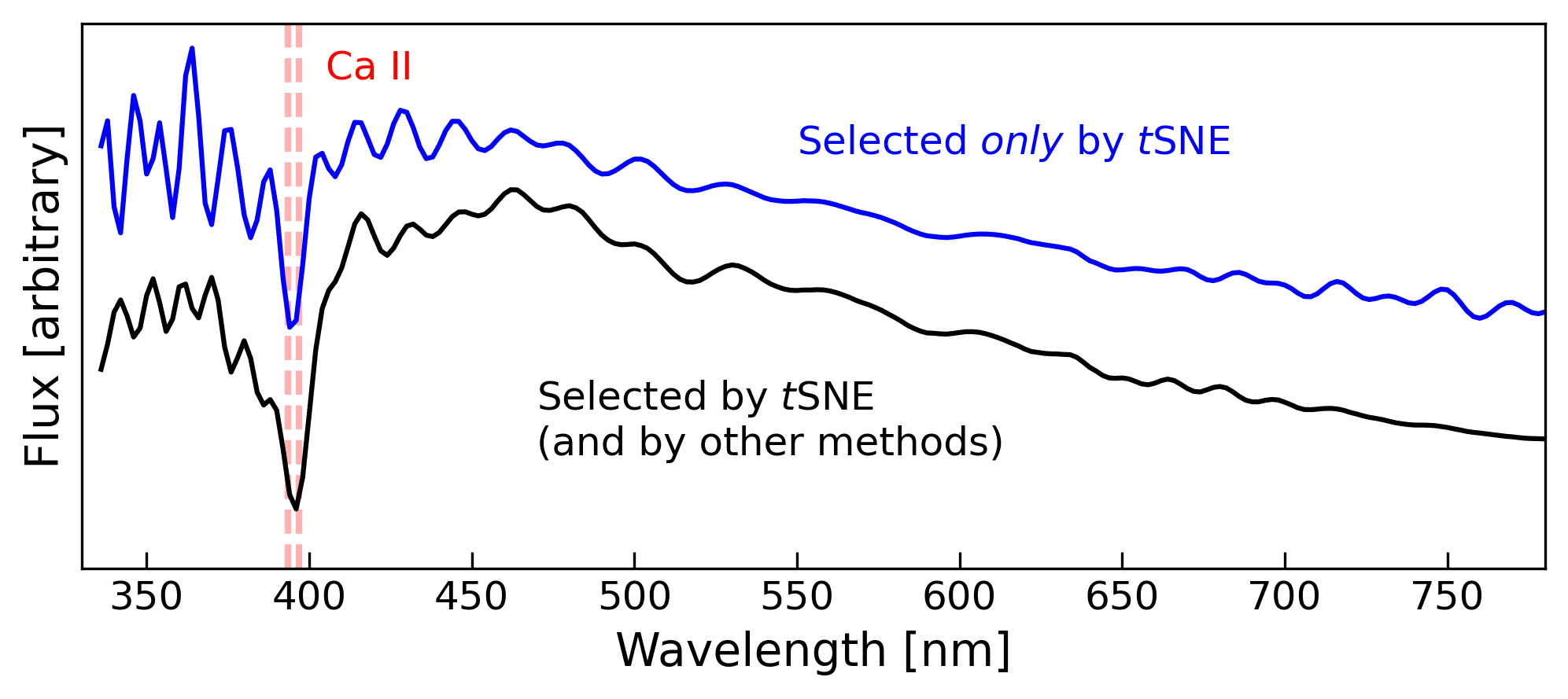}
\caption{
    Co-added \textit{Gaia} XP spectra of polluted WD candidates selected by $t$SNE that have not been observed at higher resolution (`unknown').
    The upper blue line stacks the 39 candidates which are selected only by $t$SNE.
    The lower black line stacks all 547 `unknown' candidates selected by $t$SNE, most of which are also selected by at least one other method (see Section~\ref{sec:comparison}).
    The two spectra are offset arbitrarily for visualisation purposes.
    The clear Ca~\textsc{ii} feature just below $400~\text{nm}$ suggests that many of these candidates are genuine polluted WDs.
}
\label{fig:coaddedspectrum}
\end{figure*}

With the members of each island definitively ascertained, we find that the `cool' island contains 478 sources, of which 97 are known polluted WDs and 11 are known non-polluted WDs.
The `warm' island contains 229 sources, of which 50 are known polluted WDs and 2 are known non-polluted WDs.
The remaining 547 `unknown' objects (across both islands) do not appear in any of the spectroscopic databases mentioned in Section~\ref{sec:evaluation}, but given that these objects have been clustered together based on similar features in their XP spectra, it is likely that a large proportion are as-yet-unidentified polluted WDs.
39 of these sources are not selected by any of the other methods compared later in Section~\ref{sec:comparison}, and are selected only by $t$SNE.
In the co-added XP spectrum of these `unknown' sources (Fig.~\ref{fig:coaddedspectrum}), there is a clear Ca~\textsc{ii} feature, suggesting that a significant fraction of these candidates are indeed polluted WDs.

Of the 13 `known non-polluted' objects seemingly erroneously located on the two polluted islands, most may in fact be polluted after all.
Eight are classified as DC or DA based on spectra whose wavelength coverage does not include the diagnostic Ca~\textsc{ii} features at $3933$ and $3968~\text{\AA}$.
Two others rely on very low-quality ($\mathrm{S/N}\lesssim2$) SDSS spectra classified tentatively as DCs by the data-driven pipeline of \citet{vincent23}.
As such, ten of the 13 putative false positives might not be false positives at all, though the remaining three objects appear to be genuine DCs that have been erroneously selected here.

Running UMAP on the sample (Fig.~\ref{fig:tsneclustering}(b)) gives a qualitatively similar embedding to that of \citet[cf.\ their fig.~2]{kao24}\footnote{
    The same hyperparameters were used as in the cited work: $\mathtt{n\_neighbours}=25$ and $\mathtt{min\_dist}=0.05$.
}.
The WDs from our `cool' island appear in a similar location to those selected in their work; indeed 435 sources are common to both.
However, the WDs from our `warm' island are deeply entrenched in the diagonal feature of the UMAP embedding, and could therefore not have been isolated using UMAP.

\subsection{Other features of the \textit{t}SNE embedding}

Other than polluted WDs, the $t$SNE embedding also locates other types of source, in specific locations in the embedding.
The spectral classifications in this subsection are from the \textit{Gaia}-SDSS spectroscopic sample \citep{gentilefusillo21}.

\begin{figure}
\centering
\includegraphics[width=\columnwidth]{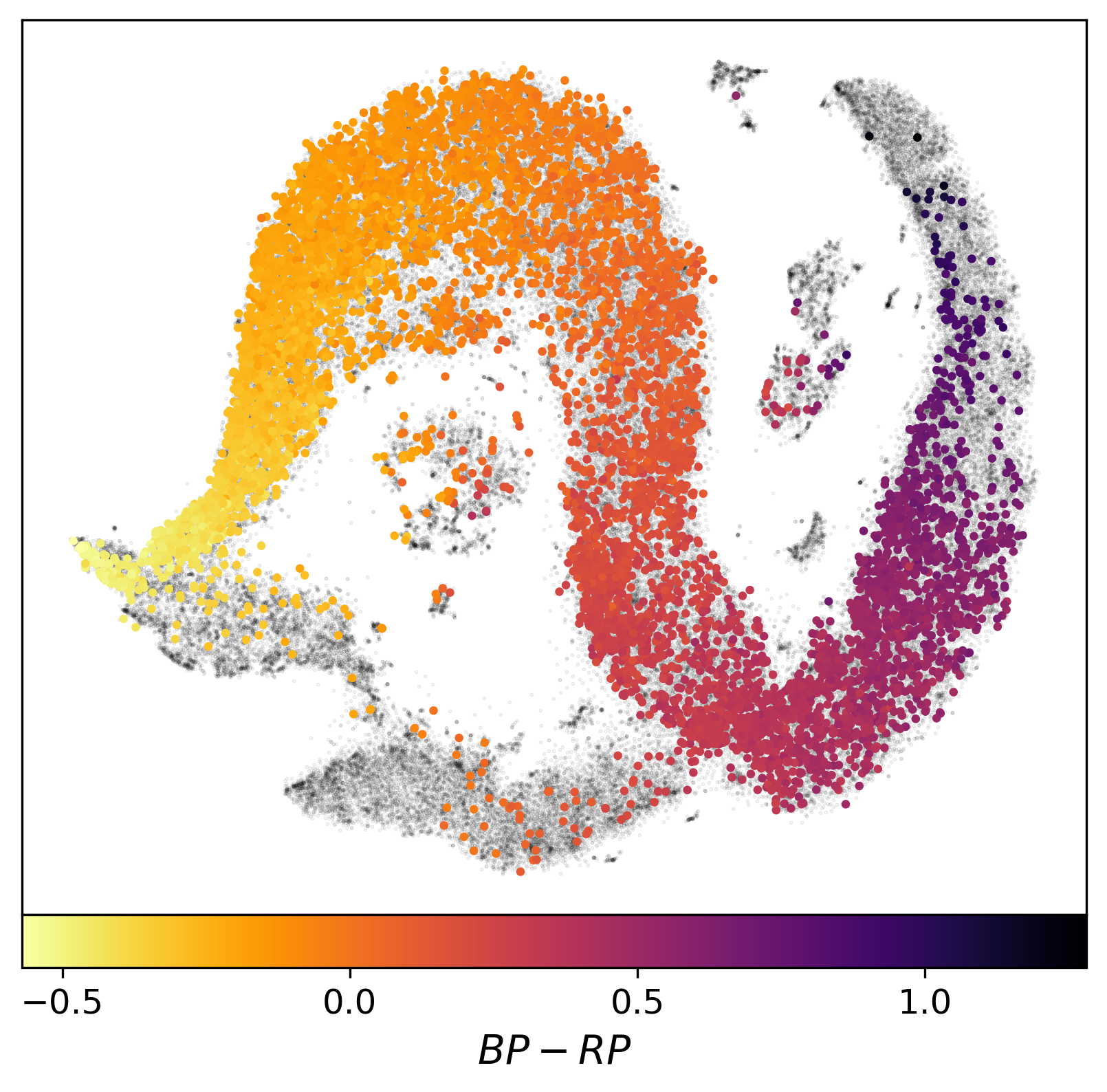}
\caption{
    Embedding and $\mathit{BP}-\mathit{RP}$ colour of spectroscopically confirmed DA WDs.
    These WDs populate the `N'-shaped sequence, with a temperature trend: the hottest DAs appear at the left-hand end; the coolest at the right.
}
\label{fig:DAs}
\end{figure}

The spectra of DA WDs are dominated by Balmer features due to atmospheric hydrogen.
Plotting their $\mathit{BP}-\mathit{RP}$ colours in the embedding (Fig.~\ref{fig:DAs}), we see that these DAs populate the `N'-shaped sequence, in decreasing temperature order from left to right.
No doubt the tilt of the black-body continuum is reflected in the XP coefficients, and hence in the embedding.
\citet{byrne24c} identify a similar trend in their application of this method to DESI EDR WD spectra, in which the primary feature of their embedding is a sequence of DAs arranged in temperature order.

\begin{figure}
\centering
\includegraphics[width=\columnwidth]{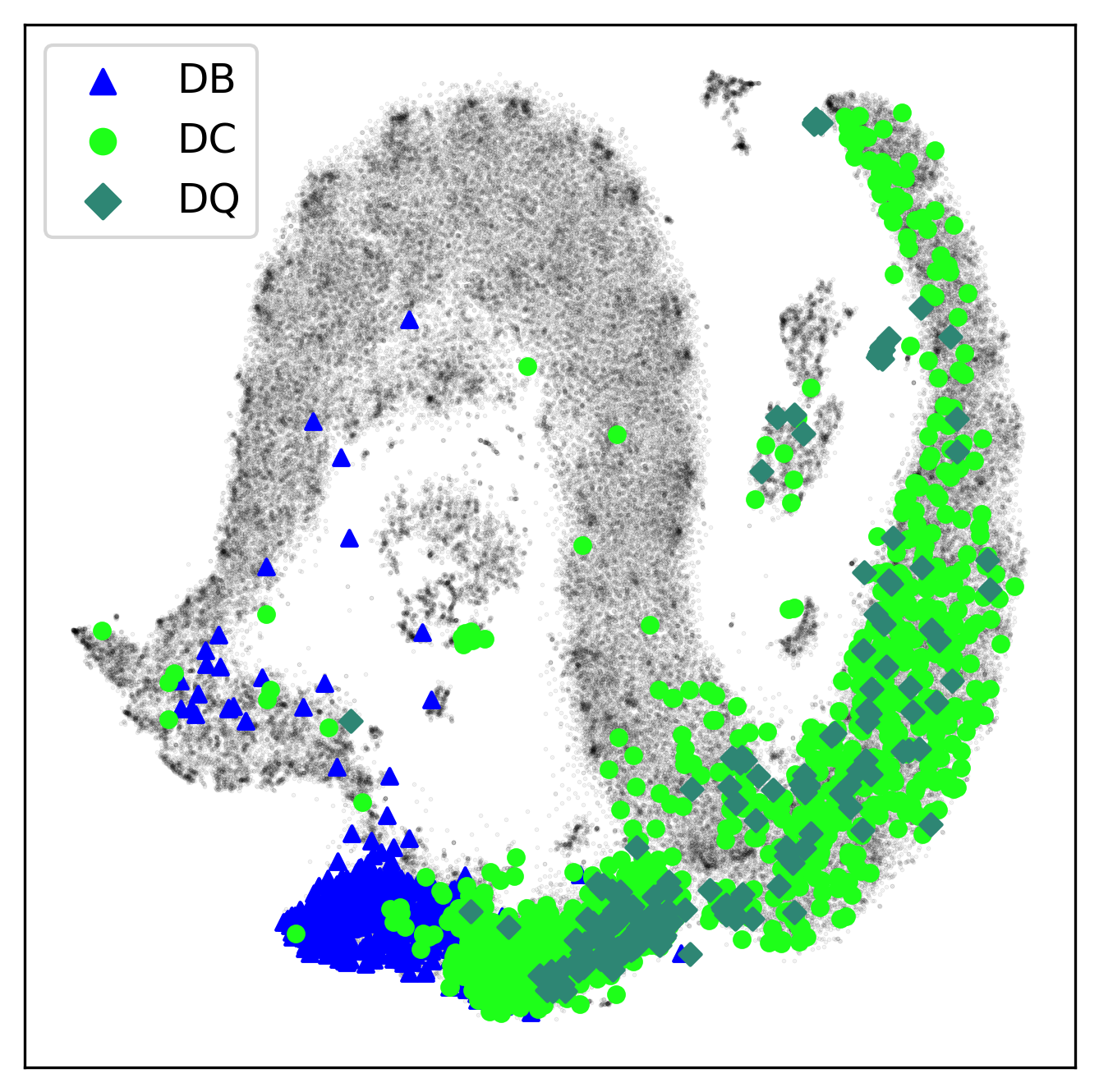}
\caption{
    Locations of spectral classes DB, DC, and DQ in the $t$SNE embedding.
    The lower island is dominated by these spectral classes, as is the `cooler' end of the `N'-shaped DA sequence.
    A smaller number of these WDs are also scattered elsewhere in the embedding.
}
\label{fig:DBCQ}
\end{figure}

The spectral classes DB, DQ, and DC correspond respectively to He~\textsc{i} features, C$_2$ Swan bands, and no features at all.
These three classes occupy the large lower island of the embedding (Fig.~\ref{fig:DBCQ}), where there is a sequence from DBs, through DCs, to DQs.
There are also large numbers of DCs and DQs along the cooler end of the `N'-shaped DA sequence.

Some aspects of the embedding mirror the principles of WD spectral evolution.
Consider a DA: it begins its life very warm, before gradually cooling over time; such a WD could be thought of as `travelling' along the `N'-shaped DA sequence.
Once the temperature cools below about $5\,500\,\text{K}$, hydrogen is no longer excited and the Balmer features gradually fade into the featureless spectrum of a DC; indeed DCs occupy the `cool end' of this sequence.
A weaker trend is seen in the helium-dominated sequence, wherein DBs transition into DQs (if carbon-rich and low-mass) or DCs \citep[e.g.][]{bedard24}.

\begin{figure}
\centering
\includegraphics[width=\columnwidth]{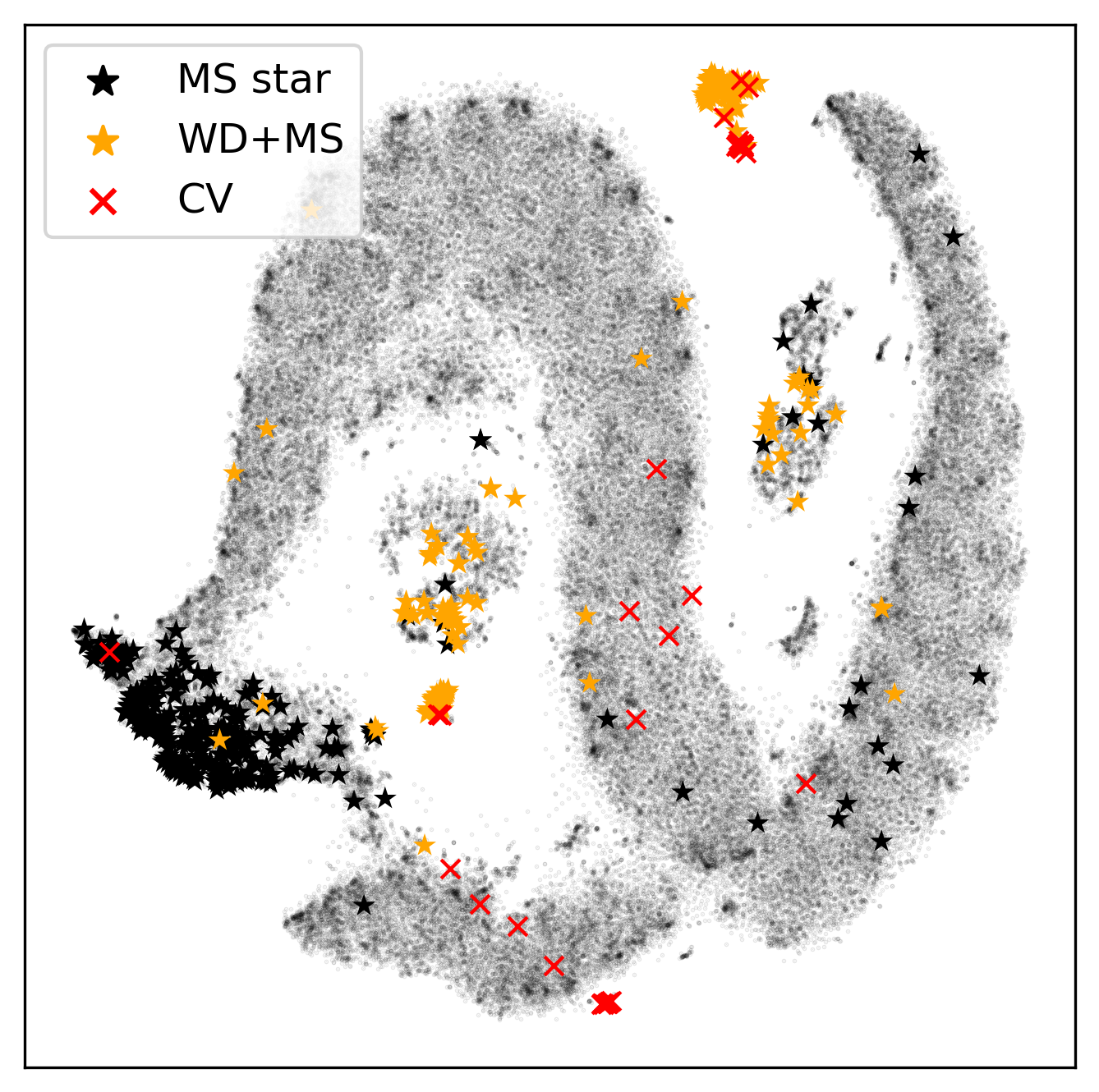}
\caption{
    Locations of MS stars, WD+MS binaries, and CVs in the $t$SNE embedding.
    MS stars are found in a trail to the left of the embedding; WD+MS binaries are found in several regions of the embedding; CVs are mostly found in an island to the top right and bottom centre.
}
\label{fig:mswdmscv}
\end{figure}

Finally, we show in Fig.~\ref{fig:mswdmscv} the locations of MS stars (erroneously included in the sample), as well as WD+MS binaries and CVs.
MS stars occupy mostly the spur on the left of the embedding.
WD+MS binaries are found in several locations in the embedding: an island in the top right; an island below left of centre; an island above right of centre.
CVs are found primarily in two places: just below the WD+MS island in the top right; on a tight island at the very bottom of the embedding.
The reasons for the identification of these spectral classes is beyond the scope of this work; suffice to say that the XP spectra of these types of object are sufficiently distinctive that they can be roughly isolated using $t$SNE.

\subsection{Comparison of methods}
\label{sec:comparison}

Here we compare the ability of five data-driven methods, including $t$SNE, in identifying polluted WDs from the \textit{Gaia} XP spectra.

\citet{garciazamora25} apply random forests, listing the classifications of $78~920$ WDs within $500\,\text{pc}$, classifying 785 as polluted (DZ).
\citet{vincent24} use gradient tree boosting to classify a larger sample of $100~886$ WD candidates, classifying $1~272$ DZs.
\citet{kao24} obtain $465$ polluted WD candidates from a sample of $96~134$ using UMAP; although this sample is not publicly available, it was acquired by personal communication.
We were able to approximately reproduce the SOM-based selection of polluted WDs by \citet{perezcouto24} identifying 451 WDs in two neurons containing a high number of known DZs (the original work reports 467).
Finally, using $t$SNE, we identify 707 polluted WD candidates across the two islands identified above (see Section~\ref{sec:results/tsne}).

We list the numbers of `known polluted', `known non-polluted', and `unknown' sources (see Section~\ref{sec:evaluation}) in each selection in Table~\ref{tab:evaluation}.
An ideal method would identify a high proportion of known polluted WDs (effectively the true positive rate) and a low proportion of known non-polluted WDs (false positive rate), and ideally a large number of sources overall.
The performance of the supervised methods is likely exaggerated, as many of the `known (non)-polluted' WDs are present in their respective training sets (see Section~\ref{sec:evaluation}).

\begin{table}
\caption{
    Pollution status of WDs classified as polluted based on their \textit{Gaia} XP spectra, according to five methods: random forests, gradient tree boosting, SOMs, UMAP, and $t$SNE.
}
\label{tab:evaluation}
 
\begin{tabular}{lccccc}
    \hline
    & RF* & GTB* & SOM$\dagger$ & UMAP & $t$SNE \\
    \hline
    Known polluted & 419 & 350 & 51 & 96 & 147\\
    & (53\%) & (28\%) & (11\%) & (21\%) & (21\%) \\
    Known non-poll. & 2 & 38 & 14 & 12 & 13 \\
    & (0.3\%) & (3.0\%) & (3.1\%) & (2.6\%) & (1.8\%) \\
    Unknown & 364 & 884 & 386 & 357 & 547 \\
    & (46\%) & (69\%) & (86\%) & (77\%) & (77\%) \\
    Total & 785 & 1~272 & 451 & 465 & 707 \\
    \hline
 \end{tabular}
*~Supervised method: performance likely overestimated (see Section~\ref{sec:evaluation}).\\
$\dagger$ Based on an approximate reproduction of \citet{perezcouto24}.
\end{table}

\begin{figure*}
\centering
\includegraphics[width=\textwidth]{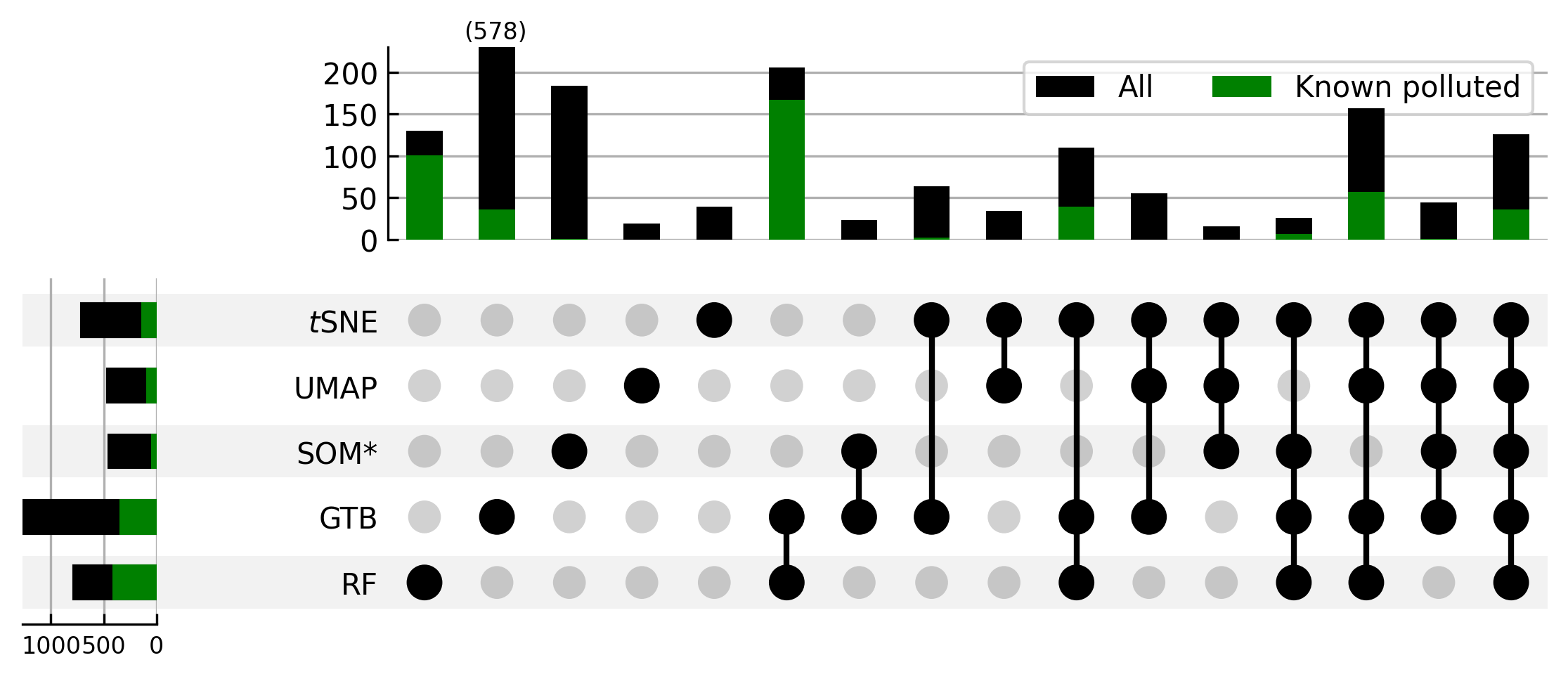}
*Based on an approximate reproduction of \citet{perezcouto24}.
\caption{
    UpSet plot \citep{lex14} showing the sources selected by various combinations of methods.
    These plots show the same information as a five-set Venn diagram, but in a less cluttered manner.
    The lower left panel shows the total number of sources identified as polluted by each method, as well as the subset of those that are known polluted WDs (these data are also given in Table~\ref{tab:evaluation}).
    The rest of the plot shows the numbers of candidates selected by only one method (first five columns), and by various combinations of methods (shown by the connectors).
    For example, 130 sources are selected only by random forests; 64 are selected by $t$SNE and gradient tree boosting and none of the other three methods; 126 are selected by all five methods (rightmost column).
    For visualization purposes, only subsets with at least 15 sources are shown, and the second column (candidates selected only by gradient tree boosting) is clipped.
    A large number of candidates are selected only by one or two of the methods, even among the known polluted WDs.
    39 unknown candidates are selected by $t$SNE and no other method.
}
\label{fig:overlap}
\end{figure*}

Random forests are by far the most successful in terms of true positive rate (53\%) and false positive rate (0.3\%).
Furthermore, \citet{garciazamora25} quote an impressive precision (fraction of selected objects which are selected correctly) of 95\% for DZs.
As such, a large proportion of the 364 `unknown' sources selected by random forests are expected to be genuine polluted WDs, though perhaps not 95\%: `known' objects are naturally biased towards objects which are nearer, have higher-quality data, and are thus more likely to be classified correctly.

Gradient tree boosting, another supervised method, outperforms the three unsupervised methods (SOMs, UMAP, $t$SNE) in identifying polluted WD candidates.
On a similar sample size ($\approx 100~000$), this method correctly recovers 350 known polluted WDs, while maintaining a low false positive rate ($3.0\%$).
Again, however, many of the known polluted WDs are present in the training set used by \citet{vincent24}, so a smaller proportion of the unknown sources might turn out to be genuine polluted WDs.

Among the unsupervised methods, UMAP and $t$SNE have similar true positive rates (21\%), though $t$SNE has a slightly better false positive rate (1.8\% vs.\ 2.6\%).
Additionally, $t$SNE proposes a larger number of polluted candidates than UMAP (707 vs.\ 465); follow-up high-resolution spectroscopic campaigns would thus be expected to return over 50\% more confirmed polluted WDs from the 547 unknown candidates selected by $t$SNE than the 357 selected by UMAP.
Finally, we note that SOMs show a lower true positive (11\%) and higher false positive rate (3.1\%) than any of the other methods.
However, we show that it can be a useful complement to other methods in Section~\ref{sec:results/overlap}.

\subsection{Overlap between polluted candidate selections} \label{sec:results/overlap}

It might be expected that each method would `learn' to identify metal-polluted XP spectra in similar ways, and hence that the candidates selected by the different methods would significantly overlap.
However, a significant number of candidates are selected only by one or two of the five methods explored here, as shown in Fig~\ref{fig:overlap}.
Only 126 candidates are selected by all five.
Even among the known polluted WDs, every method misses a significant number which were selected by at least one other method, and conversely each method selects some known polluted WDs which are not selected by any other method.

These discrepancies suggest that different methods pick up on different features in the data to identify polluted WDs.
To obtain a large polluted candidate list, evidently the best strategy would be to apply all five methods to the dataset and take the union of the methods' selections.

\section{Discussion} \label{sec:discussion}

The previous section demonstrates the use of $t$SNE in selecting candidate polluted WDs, and compares with several other data-driven methods in achieving this goal.
We begin this section by justifying the use of \textit{Gaia} XP spectra in identifying polluted candidates.
We then attempt to explain why $t$SNE isolates two islands of polluted WD candidates, rather than just one.
We then compare the $t$SNE and UMAP embeddings and discuss their differing behaviour.
Finally, we provide recommendations as to which of the methods compared here are best suited to different problems.

\subsection{Trustworthiness of \textit{Gaia} XP spectra in revealing pollution}

Given the very low resolution of the XP spectra ($R\sim 70$), it is reasonable to question the validity of methods based on these spectra to select polluted WD candidates.
The alternative is to suppose that the methods presented here merely select fortuitously-aligned noise, rather than genuine features.
We briefly justify here that these spectra are indeed useful in identifying polluted WDs.

The ultimate justification is empirical: the vast majority of sources identified as polluted WDs based on XP spectra do indeed turn out to be so, when observed at higher resolution.
A preliminary success rate of 99 per cent is reported on spectroscopic follow-up survey of the sample identified by UMAP \citep{kao24}.
We report a similarly high preliminary success rate for the $t$SNE-identified polluted candidates by cross-matching to DESI DR1 \citep{desidr1} and briefly visually inspecting the higher-resolution spectra therein.

\subsection{Why two groups of polluted WDs?}
\label{sec:whytwo}

\begin{figure*}
\centering
\includegraphics[width=\textwidth]{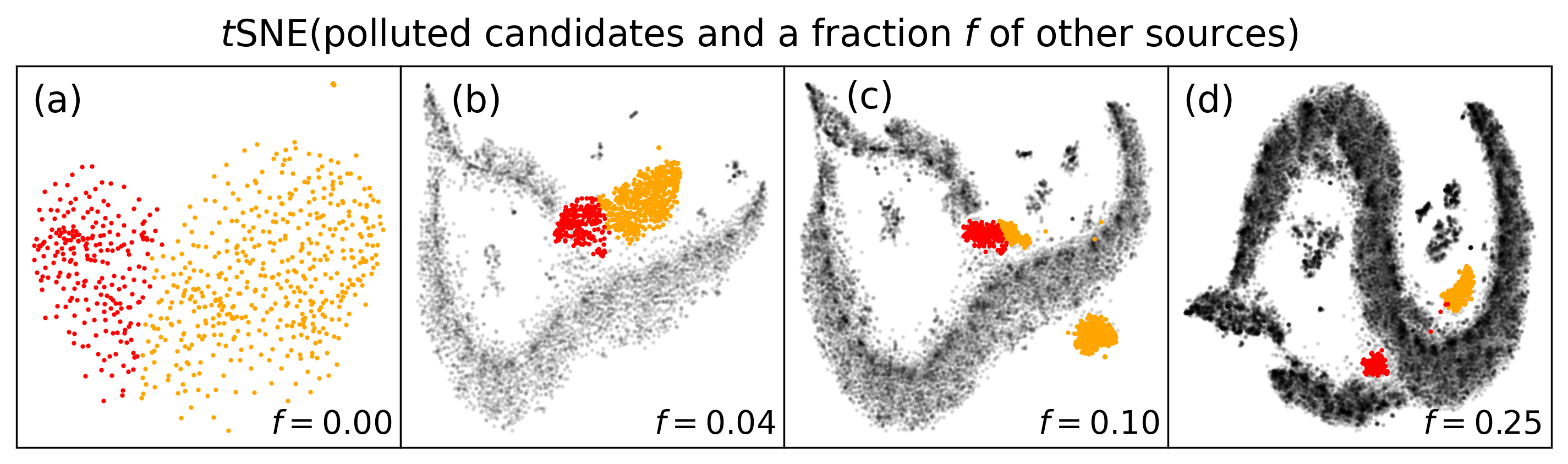}
\caption{
    $t$SNE embeddings of (a) just the polluted candidates, (b-d) the polluted candidates as well as a fraction $f$ of the rest of the sample.
    When just the polluted candidates are embedded, they do not form two distinct clusters.
    As more other objects are included, the island gradually splits into two.
}
\label{fig:whytwoislands}
\end{figure*}

The $t$SNE embedding contains two islands populated by a large portion of known polluted WDs (Fig.~\ref{fig:tsneembedding}).
One might instinctively suspect that these two islands represent two distinct populations, but they turn out to correspond to a continuum, bisected in the $t$SNE embedding by the `N'-shaped DA sequence.
If $t$SNE is run on \textit{only} the spectra on these two islands, the result is a single cluster (Fig.~\ref{fig:whytwoislands}(a)), implying that there is no natural split between the two populations.
Only when some spectra from the rest of the dataset are added to the sample do the polluted islands split into two (Fig.~\ref{fig:whytwoislands}(b-d)).

This is likely a symptom of dimensionality reduction's inability to faithfully represent the structure of the 110-dimensional dataset in 2D.
Perhaps the polluted WDs in some sense `bridge' over the DA sequence in the high-dimensional data space, and when projecting into 2D the optimal solution (in terms of similarity distributions; see Section~\ref{sec:tsne}) is to split the polluted WDs into two groups.

\begin{figure}
\centering
\includegraphics[width=\columnwidth]{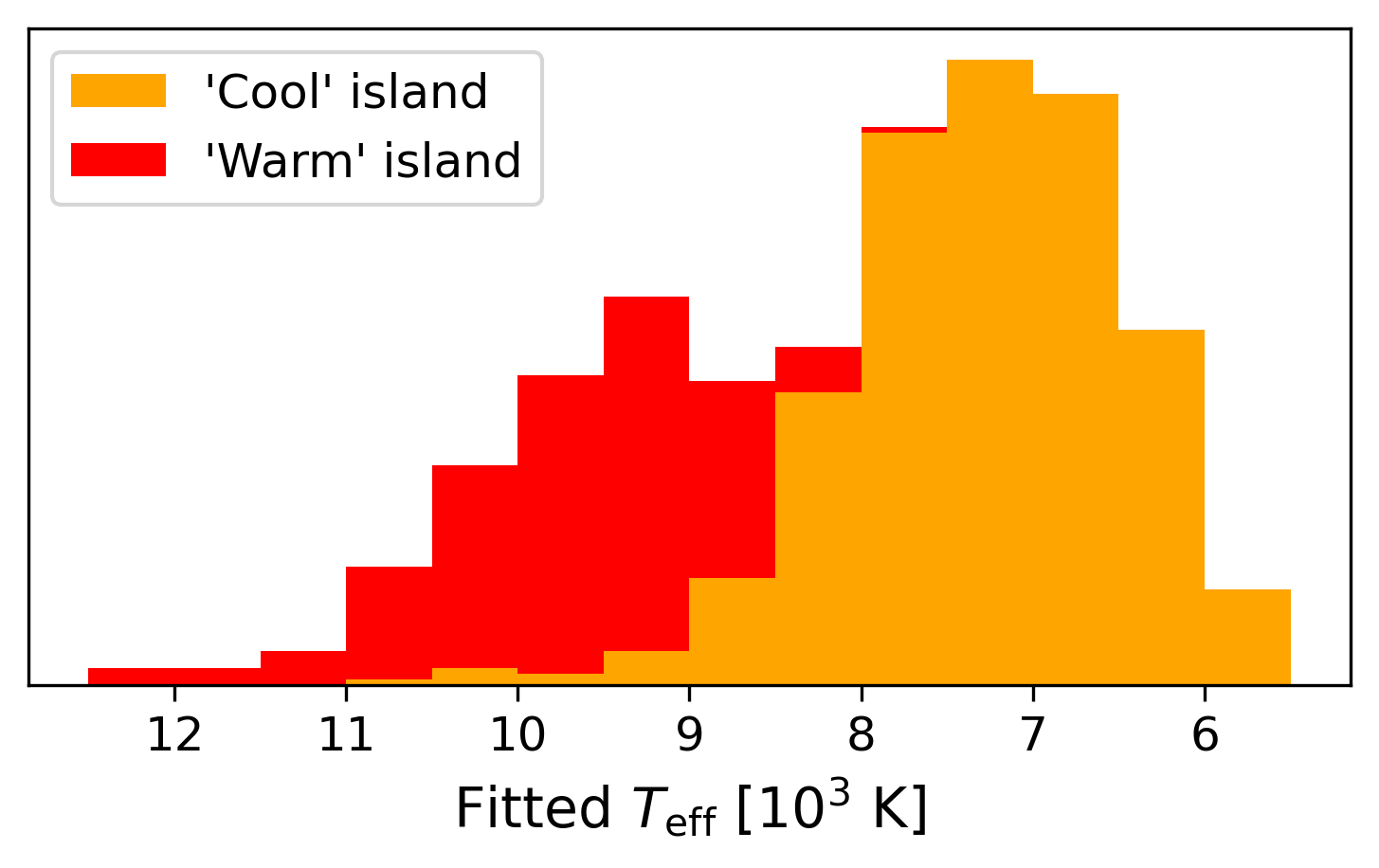}
\caption{
    Stacked histogram of estimated effective temperatures of sources selected by $t$SNE on the `cool' and `warm' islands.
    The temperatures here are of the H-atmosphere WD atmosphere model that best fit the sources' de-reddened photometry \citep{gentilefusillo21}; the temperature axis is inverted to follow the embeddings in Fig.~\ref{fig:whytwoislands}.
    Although the temperature distributions are systematically different, there is a significant overlap.
}
\label{fig:tempdistr}
\end{figure}

The estimated temperature distributions of the two polluted islands (Fig.~\ref{fig:tempdistr}) corroborate this interpretation.
While the `warm' sources are on average warmer than the the `cool' sources (hence our choice of group names), they do not clearly form two distinct distributions.
Additionally, the temperatures of the `warm' sources are similar to those of the DAs near to it in the embedding; likewise the `cool' sources.
$t$SNE empirically appears to have prioritized the co-location of sources of similar temperatures nearby, over the co-location of all polluted WDs, which are therefore split into two groups.

\subsection{Comparing \textit{t}SNE and UMAP}
\label{sec:tsnevumap}

Of the two sets of polluted WDs identified by $t$SNE, only the `cool' island corresponds to those identified using UMAP \citep[see also Fig.~\ref{fig:tsneclustering}]{kao24}.
The `warm' island is embedded by UMAP among non-polluted WDs (Fig.~\ref{fig:tsneclustering}(b)).
The two methods have similar true and false positive rates (Table~\ref{tab:evaluation}), but $t$SNE selects an overall larger number of candidates, making it a preferable method for this use case.

We suspect that $t$SNE's advantage over UMAP in this case is its deprioritization of global structure.
UMAP is known to preserve global dataset structure better than $t$SNE \citep{mcinnes18, fotopoulou24}, but for the purpose of selecting members of a rare class from a large dataset, this is not of particular importance.
Of greater importance here is the preservation of \textit{local} structure, so that as many members as possible of the rare class can be co-identified.
In prioritizing global structure, UMAP appears to sacrifice somewhat the co-identification of polluted WDs.
Another manifestation of this is the `N'-shaped DA sequence of the $t$SNE embedding, corresponding to the `U'-shaped sequence in the UMAP embedding.
As shown in Fig.~\ref{fig:DAs} (as well as fig.~2 of \citealt{kao24}), these correspond to temperature sequences, with the hottest DAs at one end and the coolest at the other.
In the UMAP embedding, this sequence is stretched out as much as possible, to preserve the distance in data space between DAs of very different temperatures.
By contrast, in the $t$SNE embedding, this sequence is coiled up into an `N' shape, as this method places less importance on distancing DAs of different temperatures.
As such, the cool end of the DA sequence (upper right of Fig.~\ref{fig:DAs}) wraps around quite close to the region occupied by intermediate-temperature DAs, even though their spectra are quite different.
Of course, global structure preservation is desirable for other use cases, such as visualizing trends across the whole dataset; for such tasks $t$SNE is known to be less appropriate than UMAP \citep{mcinnes18}.

$t$SNE is much slower than UMAP for this dataset of $\approx 100~000$ objects, taking of order $5\text{-}10\,\text{mins}$ compared to $\sim1\,\text{min}$ for UMAP; we corroborate this finding of \citet{kao24}.
$t$SNE has a time complexity of $\mathcal{O}(N \log N)$, so its use may become impractical for datasets a few orders of magnitude larger\footnote{
    Standard $t$SNE has a complexity of $\mathcal{O}(N^2)$, but the commonly-used Barnes-Hut algorithm \citep{barnes86} accelerates this to $\mathcal{O}(N\log N)$ \citep{vandermaaten13}.
}.

\subsection{Recommendations on method choice}

Selecting members of a rare class from a large dataset is a common task in Astronomy.
In this subsection we make recommendations on which of the methods mentioned above are suited to different use cases, based on insights from the search for polluted WDs from \textit{Gaia} XP spectra.

Where a significant number of high-quality training labels are available, supervised methods outperform unsupervised methods.
These labels constitute highly relevant information that supervised methods can exploit to select members of rare classes at high true positive and low false negative rates, as shown in Table~\ref{tab:evaluation}.
Care must be taken to ensure that the labels are accurate, as inaccurate training labels can have a highly detrimental effect on model performance.
Furthermore, for selecting members of very rare classes, class imbalance may become problematic, though there exist several mitigation strategies such as data augmentation and resampling \citep[see also][]{vincent25}.
In this case, the high performance of random forests and gradient tree boosting show that the benefits of training labels outweigh the obstacle of class imbalance, even without these mitigation strategies.

Among unsupervised methods, $t$SNE outperforms UMAP and SOMs in the selection of rare classes.
As discussed in Section~\ref{sec:tsnevumap}, UMAP can sacrifice the co-identification of members of rare classes in favour of preserving the global dataset structure.
SOMs -- and indeed any neural-network-based methodology -- are highly flexible, but require the tuning of a large number of hyperparameters: numbers of neurons, hidden layer sizes, learning rate, etc.
We also attempted to apply two further neural-network-based methods -- contrastive learning \citep{chen20} and disentangled representation learning \citep[e.g.][]{wang24} -- but found there to be far more hyperparameters to tune, compared to $t$SNE and UMAP\footnote{
    $t$SNE has only one main hyperparameter: the perplexity (see Section~\ref{sec:tsne}).
    UMAP has two: \texttt{n\_neighbours} and \texttt{min\_dist} (see e.g.\ \citet{kao24}, their section~3.1).
    While both methods do have further hyperparameters (such as the number of iterations to perform), the results are generally less sensitive to these.
}.

The methods examined here show complementary behaviour, each identifying candidates that were missed by the other methods (Fig.~\ref{fig:overlap}).
This suggests that a good strategy would be to apply a suite of methods and select the union of the candidates selected by each method.
However, for tasks with a different profile of class imbalance and label availability, some methods may contribute more candidates than others.
For example, a dataset with a very low number of training labels would favour the use of unsupervised methods such as $t$SNE.

\section{Conclusions} \label{sec:conclusion}

We have compared five different methods in their ability to select polluted WDs from low-resolution \textit{Gaia} XP spectra.
Among these methods we present the use of $t$SNE, which we find to outperform other unsupervised methods, identifying a population of warmer polluted candidates that was missed by UMAP, a similar technique.
39 candidates selected using $t$SNE are not selected by any other method and lack spectroscopic observations at higher resolution.
The availability of high-quality training labels is found to confer a significant advantage to supervised methods such as random forests and gradient tree boosting, especially for the highest signal-to-noise data.
We frame this work as a case study in the common task of selecting members of a rare class from a large, sparsely labelled dataset.
As Astronomy surges into the `Big Data' era, the ability to identify such interesting classes of object is crucial to a broad range of astronomical problems.

\section*{Acknowledgements}

We thank Malia Kao and Xabier P\'erez-Couto for detailed discussions and suggestions regarding their methodologies.
We thank Sarah Kane for advice on the acquisition of the \textit{Gaia} XP spectra, and Keith Hawkins for advice on normalizing them.

This work has made use of data from the European Space Agency (ESA) mission {\it Gaia} (\url{https://www.cosmos.esa.int/gaia}), processed by the {\it Gaia} Data Processing and Analysis Consortium (DPAC, \url{https://www.cosmos.esa.int/web/gaia/dpac/consortium}). Funding for the DPAC has been provided by national institutions, in particular the institutions participating in the {\it Gaia} Multilateral Agreement.
We have also made use of the Python package GaiaXPy, developed and maintained by members of the Gaia DPAC, and in particular, Coordination Unit 5 (CU5), and the Data Processing Centre located at the Institute of Astronomy, Cambridge, UK (DPCI).

We are grateful to the Leibniz-Institute for Astrophysics Potsdam (AIP) for hosting the GAIA@AIP service, which greatly streamlined the collection of the \textit{Gaia} data used in this work.

This research has made use of the VizieR catalogue access tool, CDS, Strasbourg, France (DOI: 10.26093/cds/vizier).
The original description of the VizieR service was published in \citet{ochsenbein00}.

In addition to Python packages referenced in the text, we also acknowledge the use of \textsc{Astropy} \citep{astropy1, astropy2, astropy3}, \textsc{bokeh} \citep{bokeh}, \textsc{Matplotlib} \citep{matplotlib}, \textsc{NumPy} \citep{numpy}, \textsc{pandas} \citep{pandas1, pandas2}, and \textsc{UpSetPlot}\footnote{\url{https://github.com/jnothman/UpSetPlot}}.

LKR is supported by the international Gemini Observatory, a program of NSF NOIRLab, which is managed by the Association of Universities for Research in Astronomy (AURA) under a cooperative agreement with the U.S. National Science Foundation, on behalf of the Gemini partnership of Argentina, Brazil, Canada, Chile, the Republic of Korea, and the United States of America.

Finally, we gratefully acknowledge two anonymous reviewers, whose numerous comments and suggestions greatly improved this work.

\section*{Data Availability}

The classifications of \citet{garciazamora25}\footnote{
    \url{https://cdsarc.cds.unistra.fr/viz-bin/cat/J/A+A/699/A3}
} and \citet{vincent24}\footnote{
    \url{https://cdsarc.cds.unistra.fr/viz-bin/cat/J/A+A/682/A5}
} are publicly available on the VizieR platform \citep{ochsenbein00}.
The list of polluted WD candidates selected in \citet{kao24} is not publicly available, but the embeddings are available as a VizieR catalogue\footnote{
    \url{https://cdsarc.cds.unistra.fr/viz-bin/cat/J/ApJ/970/181}
}.
All other data used in this work, including the \textit{Gaia} XP spectra, are publicly available.

Scripts used to download and process data are available at
\url{https://github.com/xbyrne/wd_xp_methods}.



\bibliographystyle{rasti} 
\bibliography{references.bib} 




\appendix


\bsp	
\label{lastpage}
\end{document}